\documentclass[aps,prb,reprint,showpacs,superscriptaddress,citeautoscript]{revtex4-1}

\usepackage{color}
\usepackage{graphicx}
\usepackage{dcolumn}
\usepackage{bm}
\usepackage[colorlinks=True,citecolor=blue,linkcolor=blue,urlcolor=blue]{hyperref}
\usepackage{amssymb, amsmath}

\begin{document}



\title{Effect of electron correlations on spin excitation bandwidth in {Ba$_{0.75}$K$_{0.25}$Fe$_{2}$As$_{2}$}
  as seen via time-of-flight inelastic neutron scattering}


\author{Naoki Murai}
\email{naoki.murai@j-parc.jp}
\affiliation{Materials and Life Science Division, J-PARC Center, Japan Atomic Energy Agency, Tokai,
  Ibaraki 319-1195, Japan}
\author{Katsuhiro Suzuki}
\affiliation{Research Organization of Science and Technology, Ritsumeikan University, Kusatsu, Shiga 525-8577, Japan}
\author{Shin-ichiro Ideta}
\affiliation{UVSOR Facility, Institute for Molecular Science,
  Okazaki 444-8585, Japan}
\author{Masamichi Nakajima}
\affiliation{Department of Physics, Osaka University, Toyonaka,
  Osaka 560-0043, Japan}
\author{Kiyohisa Tanaka}
\affiliation{UVSOR Facility, Institute for Molecular Science,
  Okazaki 444-8585, Japan}
\author{Hiroaki Ikeda}
\affiliation{Department of Physics, Ritsumeikan University, Kusatsu, Shiga 525-8577, Japan}
\author{Ryoichi Kajimoto}
\affiliation{Materials and Life Science Division, J-PARC Center, Japan Atomic Energy Agency, Tokai,
  Ibaraki 319-1195, Japan}


\date{\today}

\begin{abstract}
  \indent We use inelastic neutron scattering (INS) to investigate the effect of electron correlations on
  spin dynamics in the iron-based superconductor Ba$_{0.75}$K$_{0.25}$Fe$_{2}$As$_{2}$. 
Our INS data show a spin-wave-like dispersive feature, with a zone boundary energy of 200 meV.
A first principles analysis of dynamical spin susceptibility, incorporating
the mass renormalization factor of 3, as determined by angle-resolved photoemission spectroscopy,
provides a reasonable description of the observed spin excitations. 
This analysis shows that
electron correlations in the Fe-3$d$ bands yield enhanced effective electron masses, and consequently, 
induce substantial narrowing of the spin excitation bandwidth. 
Our results highlight the importance of electron correlations in an itinerant description of
the spin excitations in iron-based superconductors. 

\end{abstract}

\maketitle
\indent Iron-based superconductors (FeSCs) represent the second class of
high-\(\it T_{c}\) materials after the first discovery of high-\(\it T_{c}\)
superconductivity in cuprate materials. Both families have similar phase
diagrams, in which superconductivity emerges in the vicinity of an
antiferromagnetically (AFM) ordered phase. This has led to the suggestion of a 
spin-fluctuation mediated pairing mechanism, which is currently considered as a
common thread for unconventional superconductivity\cite{Scalapino_2012_Rev.Mod.Phys}. \\
\indent By contrast, a comparison between the two classes of
high-\(\it T_{c}\) families shows important differences as well.
Most importantly, unlike the cuprates, the parent compounds of FeSCs are
metals with a spin-density-wave (SDW) ground state, which, analogously to 
the SDW state in Cr metal, invokes a Fermi surface (FS) nesting picture for the
origin of the magnetism. Indeed, angle-resolved photoemission spectroscopy
(ARPES) showed that all FeSCs share a similar band structure characterized by
the presence of quasi-nested FSs
\cite{Ding_2008_Europhys.Lett, Terashima_2009_PNAS, Yoshida_2011_Phys.Rev.Lett, Richard_2011_Rep.Prog.Phys},
which enhances the tendency toward 
stripe-type AFM instability. The resulting AFM order has also been
confirmed by neutron scattering\cite{delaCruz_2008_Nature,Huang_2008_Phys.Rev.Lett,Goldman_2008_Phys.Rev.B,Zhao_2008_Phys.Rev.B,Shiliang_2009_Phys.Rev.B}.
These experimental results, as well as the superconducting gap symmetry and structure of FeSCs, can be well
explained within the framework of unconventional superconductivity caused by 
spin fluctuations of itinerant electrons\cite{Kuroki_2008_Phys.Rev.Lett,Mazin_2008_Phys.Rev.Lett,Graser_2009_New.J.Phys}.
One can thus expect that FeSCs fall in the category of itinerant electron 
systems, in contrast to the case of cuprate superconductors, in which
Mott physics is more fundamentally tied to superconductivity. \\
\indent However, while the view based on itinerant electrons is 
quite successful in FeSCs, there are important reasons to expect that
strong correlation physics may play an important role. 
First, the reduced Drude spectral weight in optical conductivity
together with bad metallic behavior indicates the strongly correlated nature of FeSCs\cite{Nakajima_2014_J.Phys.Soc.Jpn,Qazilbash_2009_Nat.Phys}.  
Secondly, density functional theory (DFT) calculations often fail to correctly describe the 
ARPES spectra of these materials owing to strong renormalization of the bands around the Fermi level\cite{Yi_2017_npj.Quantum.Materials,Yoshida_2014_Front.Phys,Derondeau_2017_Sci.Rep,Tamai_2010_Phys.Rev.Lett}. 
Both features are hallmarks of correlated metals in close proximity to a Mott insulating phase.
In view of these facts, one may now pose the following important question:
To what extent are FeSCs strongly correlated?
Therefore, it is of particular interest to characterize the strength of the
electron correlations that influence the underlying electronic and magnetic structures
of FeSCs. \\
\begin{figure*}[tbh]
\begin{center}
  \includegraphics[width= 17.00000000cm]{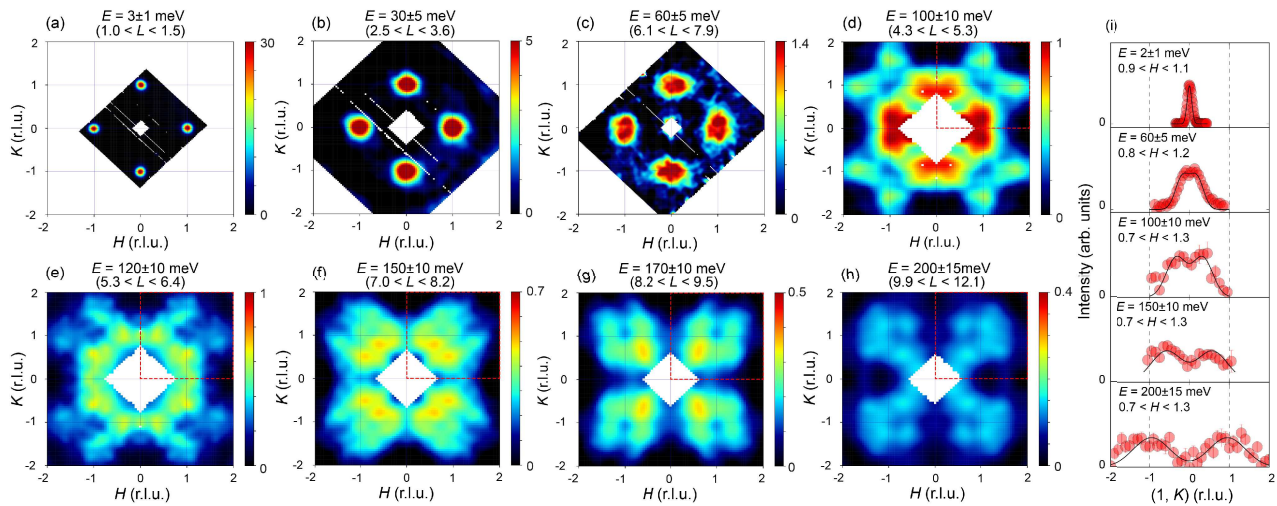}
  \caption{
    (Color online) (a)-(h) Constant-energy maps of Ba$_{0.75}$K$_{0.25}$Fe$_{2}$As$_{2}$
    in the \((H, K)\) plane at energy transfers of
    \(E = 3 \pm 1\) meV (\(E_{i} = 13.5\) meV),
    \(E = 30 \pm 5\) meV (\(E_{i} = 80.7\) meV),
    \(E = 60 \pm 5\) meV (\(E_{i} = 80.7\) meV),
    \(E = 100 \pm 10\) meV (\(E_{i} = 300\) meV), 
    \(E = 120 \pm 10\) meV (\(E_{i} = 300\) meV),
    \(E = 150 \pm 10\) meV (\(E_{i} = 300\) meV),
    \(E = 170 \pm 10\) meV (\(E_{i} = 300\) meV), and 
    \(E = 200 \pm 15\) meV (\(E_{i} = 300\) meV), respectively.
    Since the incident neutron beam was parallel to the $\it c$-axis, the value of $\it L$
      changes as a function of energy transfer. The integration ranges of $\it L$ at \({\bm Q} = (1, 0, L)\) are shown for each
      constant-energy map, respectively. The \(|{\bm Q}|\)-dependent radially symmetric background was subtracted from the raw spectra.
    For \(E > 100\) meV, data from symmetry equivalent positions in reciprocal space are averaged to improve
    statistics and folded into the area marked by the red dashed box. Constant-energy maps in (d)-(h) are obtained by
    symmetrizing the INS data in the red box.
    (i) Constant-energy cuts of the spin excitations along the
    \((1, K)\) direction at the indicated energy transfers. The black solid lines represent Gaussian fits
    to the data. 
  }   
\label{Fig1}
\end{center}
\end{figure*}
\indent Herein, we report an inelastic neutron scattering (INS) study on single crystals of hole-doped
{Ba$_{0.75}$K$_{0.25}$Fe$_{2}$As$_{2}$} that characterizes the strength of electron correlations in FeSCs
from the viewpoint of spin dynamics. 
By combining ARPES measurements and first-principles calculations, we show that the measured spin
  excitation energies are reduced by a factor of \(\sim 3\)
  compared to those obtained from the DFT-derived model. The observation of the effective (i.e., renormalized by
  electron correlations) spin excitation can easily be understood as an extension of the concept of mass renormalization to
  dynamical spin susceptibility. Our results reveal the strongly correlated nature of FeSCs beyond the DFT level, 
  which must be considered for realistic treatment of the spin dynamics in these materials.\\
\indent Single crystals of Ba$_{0.75}$K$_{0.25}$Fe$_{2}$As$_{2}$ were grown using the
FeAs-flux method, as described elsewhere\cite{Nakajima_2010_Phys.Rev.B}. 
To avoid a reaction with vaporized potassium, the starting materials, Ba, K, and FeAs, which were placed in an
aluminum crucible, were sealed in a stainless steel container\cite{Kihou_2010_J.Phys.Soc.Jpn}. 
Transport measurements show the AFM transition at \(\it T_{N}\) = 63 K, followed by the superconducting transition at
\(\it T_{c}\) = 28 K. We co-aligned 4.0 g of single crystals of Ba$_{0.75}$K$_{0.25}$Fe$_{2}$As$_{2}$ with a mosaic spread of 5$^\circ$.
INS measurements were performed using the 4SEASONS time-of-flight (TOF) chopper spectrometer
at Japan Proton Accelerator Research Complex (J-PARC)\cite{Kajimoto_2011_J.Phys.Soc.Jpn}.
Taking advantage of the repetition rate multiplication (RRM) technique for pulsed neutron
sources, a set of incident neutron energies ($\it E_{i}$'s) can be obtained 
in one experimental run, which allows for the simultaneous measurement of the low and
high energy features of an excitation spectrum\cite{Nakamura_2009_J.Phys.Soc.Jpn}. 
The measurements at 4SEASONS were performed above $\it T_{c}$ at \(T = 30 \) K\footnote{A detailed discussion of the
  so-called neutron spin-resonance and other related features observed below $\it T_{c}$ will be presented elsewhere.}
by using incident neutron energies of \(E_{i}\) = 300, 80.7, 36.8, and 13.5 meV, with corresponding energy resolutions at the
elastic line of \(\Delta E = 40, 6.5, 2.3 \) and \(0.7\) meV, respectively (full-width at half-maximum). 
The INS data were collected over a period of 3 days using a fixed sample geometry with the $\it c$-axis 
parallel to the incident neutron beam.
Data reduction of the neutron event data was performed using the {\scshape{utsusemi}} software package\cite{Inamura_2013_J.Phys.Soc.Jpn}. 
The resulting INS data were corrected for \(|{\bm Q}|\)-dependent radially symmetric
background from the sample environment\cite{Scott_2016_Phys.Rev.B} and placed on an absolute intensity scale 
(mbarn sr$^{-1}$meV$^{-1}$f.u.$^{-1}$) by using a vanadium standard\cite{Guangyong_2013_Rev.Sci.Instrum}.
In some cases, they were smoothed by convolution with a Gaussian kernel. 
Throughout this paper, we define the momentum transfer
\({\bm Q} = H{\bf a}^{*} + K{\bf b}^{*} + L{\bf c}^{*} \equiv (H, K, L)\)
in reciprocal lattice units (r.l.u.) by using the orthorhombic unit cell. 
In this notation, low-energy spin excitations associated with the stripe-type AFM order
occur at the in-plane wave vectors of \({\bm Q_{\rm AFM}} = (\pm 1, 0)\) and \((0, \pm 1)\).
ARPES experiments were performed at BL5U of the UVSOR-I\hspace{-.1em}I\hspace{-.1em}I Synchrotron by using tunable linearly polarized light of
\(h\nu = 60\) eV. Clean sample surfaces were obtained for the ARPES measurements by cleaving single crystals $in$-$situ$ in an
ultrahigh vacuum better than \(1\times10^{-8}\) Pa. The measurements were performed at \( T = 6 \) K.\\
\indent Figures~\ref{Fig1}(a)-(h) compare the two-dimensional constant energy maps of spin excitations
in the \((H, K)\) scattering plane for various energy transfers.
At low energies below 60 meV [Figs.~\ref{Fig1}(a)-(c)], spin excitations peak strongly at 
\({\bm Q_{\rm AFM}}\), which corresponds to the nesting vector between hole and electron FSs.
As the energy increases, spin excitations form transversely elongated ellipses
that lead to splitting into two branches [Figs.~\ref{Fig1}(d) and (e)]. 
At even higher energies above \(150\) meV, these excitations broaden rapidly and 
form broad circular shapes centered at the zone boundary \({\bm Q} = (\pm 1, \pm 1)\)
[Figs.~\ref{Fig1}(f)-(h)]. These dispersive features are also confirmed by the constant energy cuts along
the \((1, 0) \rightarrow (1, \pm 1)\) symmetry direction as shown in Fig.~{\ref{Fig1}}(i), where
a single commensurate peak centered at \((1, 0)\) at low energies (\(E < 60\) meV) splits into a pair of two peaks
with increasing energy, and eventually moves close to the zone boundary \({\bm Q}=(1, 1)\) at
\(E \sim 200\) meV. \\
\indent The dispersive spin-wave-like structure can be seen more clearly in the \({\bm Q}\)-\(E\) maps.
Figures~{\ref{Fig2}(a) and \ref{Fig2}(b)} compare the low- and high-energy features of spin excitations projected
along the \((1, K)\) high-symmetry direction. 
The spin excitations at low energies are seen to be steeply dispersing and
concentrated solely in the region near \({\bm Q}_{\rm AFM}\) [Fig.~{\ref{Fig2}(a)}].
As the energy increases, they disperse along the 
\((1, 0) \rightarrow (1, \pm 1)\) symmetry direction until the energy reaches \(E \sim 200\) meV
near the zone boundary [Fig.~\ref{Fig2}(b)].
The observed spin excitation bandwidth of Ba$_{0.75}$K$_{0.25}$Fe$_{2}$As$_{2}$ is similar to that reported for 
the parent and electron-doped BaFe$_{2}$As$_{2}$ systems\cite{Dai_2015_Rev.Mod.Phys,Tranquada_2014_J.Magn.Magn.Mater,Harriger_2011_Phys.Rev.B,Harriger_2012_Phys.Rev.B,Liu_2012_Nat.Phys,Luo_2013_Phys.Rev.B,Wang_2013_Nat.Commun,Li_2010_Phys.Rev.B,Tucker_2012_Phys.Rev.B}. \\
\indent To understand the INS data, we provide a first-principles analysis of the spin excitation spectrum of FeSCs
on the basis of the itinerant picture. 
As the first step, we obtain the DFT band structure of BaFe$_{2}$As$_{2}$ by using the
{\scshape{quantum espresso}} package\cite{Giannozzi_2009_J.Phys.Condens.Matter}
with the experimental lattice parameters\cite{Rotter_2008_Phys.Rev.Lett}.
Here, we adopt the generalized gradient approximation (GGA) exchange-correlation functional\cite{Perdew_1996_Phys.Rev.Lett},
and take cutoff energy of \(E_{\rm cut}\) = 40 Ry and 512 $\it k$-point mesh. Then, we construct an effective
five-orbital tight-binding model by using the maximally localized Wannier functions
(MLWFs)\cite{Marzari_1997_Phys.Rev.B,Souza_2001_Phys.Rev.B,Arash_2008_Comput.Phys.Commun} and 
the unfolding procedure developed in Ref.[\onlinecite{Suzuki_2011_Phys.Rev.B}].
The refolded band structure of the five-orbital model shows good agreement with the result of the original DFT
calculations [Fig. \ref{Fig3}(a)]. 
The effect of K substitution in Ba$_{1-x}$K$_{x}$Fe$_{2}$As$_{2}$ is treated by the rigid-band shift of
the Fermi level, as its validity has been confirmed by the ARPES study\cite{Liu_2008_Phys.Rev.Lett}.
Considering Hubbard-type interactions (i.e., the intraorbital Coulomb repulsion $\it U$,
interorbital Coulomb repulsion $\it U^{'}$, Hund's coupling $\it J$ and pair hopping $\it J^{'}$),
we obtain the dynamical spin susceptibility \(\hat{\chi}_s(\bm q,\omega)\) within the random phase approximation (RPA) as,   
\begin{align}
  \hat{\chi}_s({\bm q}, E)&=\hat{\chi}_0({\bm q}, E)[\hat{I}-\hat{S}\hat{\chi}_0({\bm q}, E)]^{-1}. 
  \label{eqchi}
\end{align}
Here, $\hat{S}$ is the corresponding interaction vertex matrix\cite{Yada_2005_J.Phys.Soc.Jpn} 
and $\hat \chi_0({\bm q}, E)$ is the irreducible susceptibility given as 
\begin{align}
  \begin{split}
    \chi^{l_1,l_2,l_3,l_4}_0(&{\bm q}, E)=\sum_{\bm k}\sum_{n,m}\frac{f(\varepsilon^n_{\bm k+\bm q})-f(\varepsilon^m_{\bm k})}{E+i\delta-\varepsilon^n_{\bm k+\bm q}+\varepsilon^m_{\bm k}}\\
    &\times U_{l_1,n}(\bm k+\bm q)U_{l_4,m}(\bm k)U^{\dagger}_{m,l_2}(\bm k)U^{\dagger}_{n,l_3}(\bm k+\bm q), 
    \label{eqchi0}
  \end{split}
\end{align}
where $f(\varepsilon)$, \(\varepsilon^m_{\bm k}\) and \(U_{1,n}(\bm k)\) are, respectively,
the Fermi distribution function, energy dispersion, and elements of the unitary matrices from the orbital to the band basis. 
For the results shown below, we set $U=1.0$~eV, $U'=U-2J$, $J=J'=U/8$, \(T = 1.5\times10^{-2}\) eV,
\(256 \times 256 \times 1\) $\it k$-point meshes, and smearing factor of \(\delta=1.6\times10^{-2}\) eV. \\
\indent Figure \ref{Fig3}(b) shows a contour plot of the calculated \(\chi_{s}({\bm q}, E)\)
along the high-symmetry directions. The spin excitations are markedly different in the
two directions \((1, 0)\rightarrow(1, 1)\) and \((1, 0)\rightarrow(0, 0)\), which is
consistent with the transversely elongated spin excitations seen in Fig.~\ref{Fig1}. 
Such highly anisotropic spin excitations are typical of FeSCs\cite{Dai_2015_Rev.Mod.Phys,Tranquada_2014_J.Magn.Magn.Mater,Zhao_2009_Nat.Phys,Harriger_2011_Phys.Rev.B,Harriger_2012_Phys.Rev.B,Liu_2012_Nat.Phys,Luo_2013_Phys.Rev.B,Wang_2013_Nat.Commun,Ewings_2011_Phys.Rev.B,Tucker_2012_Phys.Rev.B}. 
However, while our RPA calculation reproduces the dispersive spin-wave-like feature along
the high-symmetry directions, it apparently overestimates the energy scale of the excitations.
Along the \((1, 0) \rightarrow (1, 1)\) direction, the theoretical spin excitation peak extends nearly to
\(E \sim 600\) meV, which is larger by a factor of \(\sim 3\) than the experimental data. \\
\begin{figure}[tb]
\begin{center}
  \includegraphics[width= 8.7800cm]{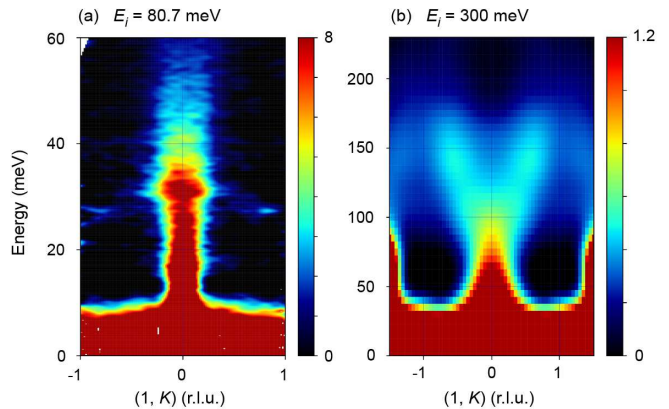}
  \caption{
    (Color online) Energy band dispersion of spin excitations along the \((1, K)\) direction  
    measured with \(E_{i} = 80.7\) and 300 meV for panels (a) and (b), respectively.
    The map in (b) is obtained by symmetrizing the raw data with respect to the \( K = 0\) mirror plane. 
    }
\label{Fig2}
\end{center}
\end{figure}
\indent This overestimation of the spin excitation energy in our RPA analysis has  
important implications for the electronic state of FeSCs.
As can be seen in Eq.~{\eqref{eqchi0}}, the electronic band structure \(\varepsilon^m_{\bm k}\) determines 
the momentum- and energy-dependent structure of the dynamical spin susceptibility.  
The discrepancy between the experimental and theoretical spin excitations, therefore,  
suggests that the actual electronic structure of Ba$_{0.75}$K$_{0.25}$Fe$_{2}$As$_{2}$ deviates from the
DFT-derived model. In principle, DFT provides a good starting point for modeling the electronic structure of the
weakly correlated regime. However, when electron correlations become sizable,
the low-energy bands near the Fermi level are heavily renormalized,
which results in a substantial effective mass (\(m^{*}\)) enhancement (or equivalently, bandwidth (\(W\)) narrowing)
relative to DFT calculations. Spectroscopic probes such as ARPES can provide direct information about the real
single-particle spectra of the correlated materials, which cannot be accurately captured by DFT. \\
\begin{figure}[tb]
\begin{center}
  \includegraphics[clip,width= 8.5000cm]{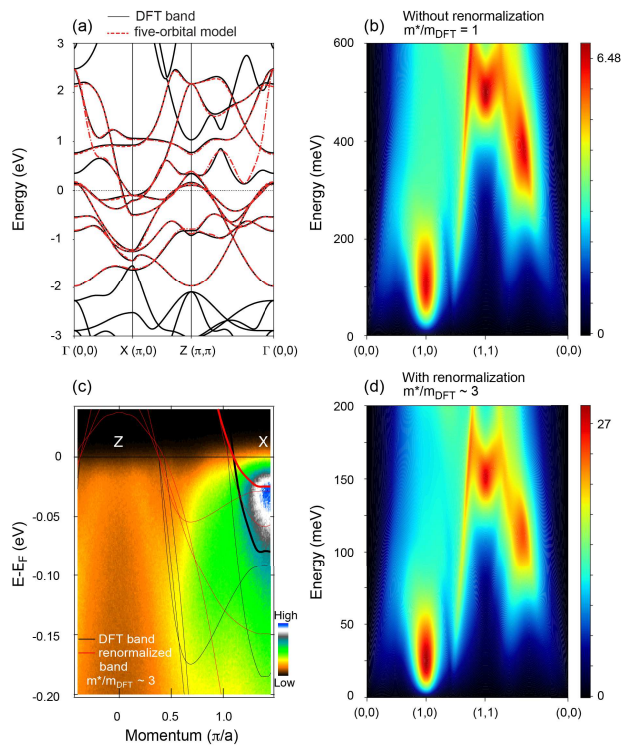}
  \caption{
    (Color online) (a) Electronic band structure of BaFe$_{2}$As$_{2}$ along high-symmetry directions. 
    Black solid lines denote the DFT band structure, whereas red dashed dotted lines denote 
    the effective five-orbital model obtained using MLWFs. 
    (b) Energy band dispersion of RPA dynamical spin susceptibility, \(\chi_{s}({\bm q}, E)\), 
    along high-symmetry directions. The RPA calculation was performed for the original 
    DFT-derived band structure without mass renormalization (\(m^{*}/m_{\rm DFT} = 1\)). 
     (c) Comparison between ARPES and DFT band structures along 
  the \(Z\)-\(X\) direction. The black and red lines denote the original \((m^{*}/m_{\rm DFT} = 1)\) and the
  renormalized (\(m^{*}/m_{\rm DFT} \sim 3\)) DFT bands, respectively.
  The $d_{xz/yz}$ electron bands at the $\it X$ point are denoted by the thick lines. 
  (d)  Energy band dispersion of RPA dynamical spin susceptibility, \(\chi_{s}({\bm q}, E)\), 
  along high-symmetry directions.
  To account for the ARPES-derived mass enhancement factor,
  the RPA calculation was performed for the renormalized (\(m^{*}/m_{\rm DFT} \sim 3\)) DFT band structure.  
  }
\label{Fig3}
\end{center}
\end{figure}
\indent To gain more insight into the experimental electronic structure, 
we performed ARPES measurements on crystals from the same batch as that used for INS measurements. 
As expected for the correlated state, mass renormalization relative to the DFT calculations
\((m^{*}/m_{\rm DFT})\), which quantifies the strength of electron correlation, was clearly observed. 
Figure~\ref{Fig3}(c) shows the spectral image of Ba$_{0.75}$K$_{0.25}$Fe$_{2}$As$_{2}$ along the
$\it Z\mathchar`-X$ high-symmetry direction overlaid with DFT bands.
The high-intensity region at the \(X\)-point corresponds to the bottom of the \(d_{xz/yz}\) electron band. 
The overall bandwidth narrowing is estimated by scaling the DFT band to fit the
\(d_{xz/yz}\) band bottom at $X$.
\footnote{A simple rescaling of the DFT bands is not enough to fully account for the ARPES data.
  We therefore used the energy position of the band bottom at the \(X\) point as an overall measure of
  the band renormalization.
  We note that while the ARPES measurements were performed below $\it T_{c}$, the overall electron bandwidth is not affected by the
  occurrence of superconductivity.}
This analysis yields \(m^{*}/m_{\rm DFT} \sim 3\), which is in reasonable agreement with estimates from the fluctuation exchange
(FLEX) approximation\cite{Ikeda2010Phys.Rev.B}. 
Interestingly, the obtained mass enhancement of \(m^{*}/m_{\rm DFT} \sim 3\) is surprisingly close to the
renormalization factor of the spin excitation bandwidth.
Such consistency between the results of INS and ARPES has not been reported before. \\
\indent Given the strong sensitivity of the spin excitations to the underlying electronic structure,
one can expect that the Fe-3$\it d$ bandwidth narrowing due to electron correlations is directly
reflected in the spin excitation energy scale.
To confirm this, we reevaluated the dynamical spin susceptibility under the scaling of the electron band energy as 
\(\varepsilon^m_{\bm k} \rightarrow \varepsilon^m_{\bm k}/z\). (Here, \(z = m^{*}/m_{\rm DFT}\) is
the ARPES-derived mass enhancement factor.)
In addition, we applied similar renormalization to the on-site interactions, temperature, and smearing factor.
As shown in Fig. \ref{Fig3}(d), this scaling reduced the spin excitation bandwidth to \(\sim 1/3\)
of its original width, yielding a broadly consistent description of the observed INS data\footnote{Here, we mention that
  the mass-renormalized spin susceptibility exhibits a peak structure at around \(E \sim 500\) meV,
  which is higher than the experimentally accessible energy range of the present study.  
  This excitation is consistent with a previous study of the 1111 system without mass
  renormalization\cite{Kariyado_2009_J.Phys.Soc.Jpn}. However, one should note that, in the RPA treatment, 
  the dynamical effects of the quasiparticle lifetime are neglected.
  Therefore, the experimental spin excitations at such high energies would largely be obscured by the
  finite quasiparticle lifetime}. 
The reasons for this consistency can easily be understood as follows. 
The dispersion of spin excitations is defined by the resonance condition in Eq.~({\ref{eqchi0}}), in which 
the denominator of irreducible susceptibility becomes zero.
If the electron band energy is renormalized as \(\varepsilon^m_{\bm k} \rightarrow \varepsilon^m_{\bm k}/z\), 
correspondingly, a spin excitation peak energy, defined as \(\varepsilon^n_{{\bm k}+{\bm q}} - \varepsilon^m_{\bm k}\),
shows similar renormalization, much like the electron bands.
This means that the concept of mass renormalization in the Fermi-liquid theory can be extended to
dynamical spin susceptibility.
A similar discussion can be used to understand the material- and doping-dependent trends of spin-excitation
bandwidth in FeSCs, as has been shown using the dynamical mean field theory (DMFT)\cite{Zhang_2014_Phys.Rev.Lett,Ding_2016_Phys.Rev.B,Man_2017_npj.Quantum.Materials}. \\ 
\indent Our results thus demonstrate that it is possible to model the spin excitations of FeSCs by incorporating
aspects of the low-energy quasiparticle renormalization that affect both single- and two-particle quantities. 
In addition, the consistency of the mass renormalization factors determined by independent
INS and ARPES measurements highlights the potential capability of INS for characterizing 
the strength of electron correlations. The observed mass renormalization \(m^{*}/m_{\rm DFT}\sim3\) is comparable to
that of typical correlated metals such as SrVO$_{3}$ \( (m^{*}/m_{\rm DFT}\sim2) \)\cite{Yoshida_2005_Phys.Rev.Lett} and
Tl$_{2}$Ba$_{2}$CuO$_{6+\delta}$ \((m^{*}/m_{\rm DFT}\sim3)\)\cite{Rourke_2010_New.J.Phys}. 
In this context, it is interesting to recall a recent first-principles study\cite{Miyake_2010_J.Phys.Soc.Jpn} 
that estimated the strength of electron correlations, \(\it U/t \ (U/W)\), by using the constrained RPA method.  
(Here, \(t\) is the nearest-neighbor hopping parameter.) This calculation yields 
\(U/t \ (U/W) = \) {8--14} (0.5--0.9) for FeSCs as an average of the five orbitals\cite{Miyake_2010_J.Phys.Soc.Jpn}, 
which is comparable to or even larger than \(U/t \ (U/W)\) = {2--7} (0.2--0.8) obtained for
cuprates\cite{Jang_2016_Sci.Rep}. Our results, taken together with these considerations, suggest that
FeSCs have stronger electron correlations than previously expected\cite{Yang_2009_Phys.Rev.B}, and more importantly, 
such a correlated electronic state is a crucial aspect that must be considered to realistically describe spin dynamics.
With recent advances in modern INS spectrometers at spallation neutron sources, it is now becoming possible to experimentally determine 
the complicated spin susceptibility arising from the correlated band structure\cite{Li_2016_Phys.Rev.Lett,Goremychkin_2018_Science}, and in the
future INS will allow us to discuss both magnetic and electronic structures on an equal footing. \\
\indent To summarize, we performed a combined INS and ARPES study on Ba$_{0.75}$K$_{0.25}$Fe$_{2}$As$_{2}$
that revealed the effects of electron correlations on spin dynamics in FeSCs. 
The measurements show, in combination with first-principles calculations, that the correlation-induced narrowing of the Fe-3$d$ bandwidth
is reflected directly in the spin-excitation bandwidth.
Our analysis of the spin excitation spectrum provides much richer information on the nature of electron correlations
than can be obtained in a conventional analysis based on the spin-only Hamiltonian.
In addition, the two independent momentum-resolved techniques used in the present study,
INS and ARPES, are closely related, and provide the same mass renormalization factors consistently.
These results highlight the potential of INS for use as a
momentum-resolved probe for determining the electronic structure of correlated electron systems. \\
\indent Neutron scattering experiments at the Materials and Life Science Experimental
Facility of J-PARC were carried out under proposals No.2015I0001, No.2016I0001, No.2016B0263 and
No.2017I0001. ARPES experiments at UVSOR-I\hspace{-.1em}I\hspace{-.1em}I Synchrotron were carried out
under proposals No.28-529, No.28-820, and No.29-530. 
This work was partly supported by Grants-in-Aid for Scientific Research (Grants No.JP15K04742, No.JP15K17709, No.JP17J06088, No.JP16H04021 and No.JP16H01081),
JSPS, Japan. We acknowledge valuable discussions with K. Nakajima and S. Ohira-Kawamura, and technical support from
W. Kambara, K. Aoyama, H. Isozaki, and K. Ikeuchi. \\
\bibliography{bibliography}

\end{document}